\documentclass{iau} 
\usepackage{graphicx}
\usepackage{amssymb}
\usepackage{natbib}

\title[Evidence of thermal conduction suppression] 
{Evidence of thermal conduction suppression in hot coronal loops: Supplementary results}

\author[Wang et al.]   
{Tongjiang Wang$^{1,2}$, Leon Ofman$^{1,2}$, Xudong Sun$^3$, Elena Provornikova$^4$
\and Joseph M. Davila$^{2}$}

\affiliation{$^1$Dept. of Physics, Catholic University of America, 620 Michigan Avenue NE, Washington, DC 20064, USA;   email: {\tt tongjiang.wang@nasa.gov} \\[\affilskip]
$^2$NASA Goddard Space Flight Center, Code 671, Greenbelt, MD 20770, USA \\[\affilskip]
$^3$W. W. Hansen Experimental Physics Laboratory, Stanford University, \\Stanford, CA 94305, USA\\[\affilskip]
$^4$ Space Science Division, Naval Research Laboratory, Washington, DC 20375, USA\\[\affilskip]}

\pubyear{2015}
\volume{xxx}  
\jname{IAU Symposium 320: Solar and Stellar Flares and Their Effects on Planets }
\editors{A.G. Kosovichev, S.L. Hawley, \& P. Heinzel, eds.}
\begin{document}

\maketitle

\begin{abstract}
Slow magnetoacoustic waves were first detected in hot ($>$6 MK) flare loops by the SOHO/SUMER
 spectrometer as Doppler shift oscillations in Fe\,{\sc xix} and Fe\,{\sc xxi} lines. 
Recently, such longitudinal waves have been found by SDO/AIA 
in the 94 and 131 \AA\ channels. \citet{wan15} reported the first AIA event revealing signatures
in agreement with a fundamental standing slow-mode wave, and found quantitative evidence
for thermal conduction suppression from the temperature and density perturbations in the hot 
loop plasma of $\gtrsim$ 9 MK. The present study extends the work of \citet{wan15} by 
using an alternative approach. We determine the polytropic index directly based on 
the polytropic assumption instead of invoking
the linear approximation. The same results are obtained as in the linear approximation,
indicating that the nonlinearity effect is negligible. We find that the flare loop cools 
slower (by a factor of 2--4) than expected from the classical Spitzer 
conductive cooling, approximately consistent with the result of conduction suppression 
obtained from the wave analysis. The modified Spitzer cooling timescales based on the
nonlocal conduction approximation are consistent with the observed, suggesting that
nonlocal conduction may account for the observed conduction suppression in this event. 
In addition, the conduction suppression mechanism predicts that larger flares may tend to
be hotter than expected by the EM-$T$ relation derived by \citet{shi02}. 

\end{abstract}

\firstsection 
\section{Introduction}
The magnetically dominated plasma of the solar corona can support propagation of various types of magnetohydrodynamics (MHD) waves. Observations of these waves allow us to determine physical parameters of coronal structures that cannot be measured directly via a technique called 
{\it coronal seismology} \citep{rob83, nak05, liuw14}. The knowledge of the appropriate 
value of the polytropic index is important for understanding the energy transport
and the relation between density, temperature and pressure in hydrodynamic and MHD models of 
the solar and stellar coronae as well as of space plasmas \citep[e.g.][]{jac11}. 
In contrast to the adiabatic index that is always 5/3 for an ideal monatomic gas (or fully ionized
ideal plasma), the polytropic index can have other values to account for the overall
contribution of the different physical processes (e.g., heating, radiative cooling, and 
thermal conduction) in the energy equation. From Hinode/EIS observations of propagating 
slow magnetoacoustic waves in a coronal loop, \citet{van11} estimated the polytropic index 
to be close to 1, and suggested that thermal conduction is the dominant heat transport mechanism 
in the warm (1--2 MK) corona.

Impulsive energy release in a closed magnetic structure (so-called confined or non-eruptive
flares) provides a natural scenario for excitation of slow magnetoacoustic waves. This phenomenon
was first discovered by the SOHO/SUMER spectrometer in hot coronal loops as Doppler shift
oscillations in the flare lines \citep[][for a review]{wan02, wan11}. These oscillations 
were mainly interpreted as fundamental standing slow modes because their phase speed is close 
to sound speed in the loop, and there is a quarter-period phase shift between the velocity
and intensity oscillations in some events \citep{wan03a, wan03b}. The initiation 
of the waves was often associated with small flares at the loop footpoint \citep{wan05}.
These waves typically show a rapid decay. Theoretical and numerical studies suggested that
the dominant dissipation mechanisms are thermal conduction \citep{ofm02, dem03}, 
compressive viscosity \citep{sig07}, and nonlinearity effect \citep{rud13}. These
hot loop oscillation events are characterized by impulsive heating followed by
a gradual cooling phase with similar features as solar flares, so also referred to as hot loop
transient events \citep[HLTEs,][]{curd04}. The HLTEs observed in multiple 
spectral lines formed at different temperatures have been used to diagnose the heating function
by a forward modeling approach \citep{tar07}. In addition, the wave periods were also used
to determine the loop magnetic field by coronal seismology \citep{wan07}.

\begin{figure}[t]
\begin{center}
 \includegraphics[width=5in]{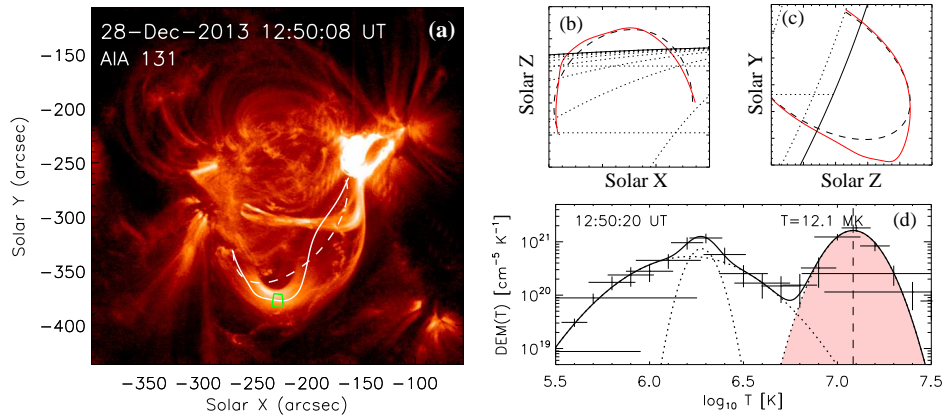} 
 \vspace*{-0.4 cm}
 \caption{A longitudinal wave event observed by SDO/AIA on 2013 December 28. (a) 131 \AA\ image. 
The solid curve indicates the oscillating loop and its 3D reconstruction by the method of
curvature radius maximization, and the dashed curve represents a fitted circular model.
(b) and (c): The 2D projections of the loop models in the XZ and YZ planes, where Z is 
the direction of the observer's line-of-sight. (d) The derived differential emission measure 
(DEM) profile ({\it crosses}) for a small region ($9^{''}$$\times$$13^{''}$) 
marked with a box in (a). A fitted triple-Gaussian model ({\it curves}) is used to isolate 
the hot loop contribution ({\it pink}) from the background.}
   \label{fig1}
\end{center}
\end{figure}

Recently, \citet{kum13, kum15} reported longitudinal wave events observed with the Solar Dynamics Observatory (SDO)/Atmospheric Imaging Assembly (AIA). The waves have similar physical 
properties as hot loop oscillations observed by SUMER \citep{wan03b, wan07}, however, 
they bore the feature of bouncing back and forth in the heated loop, 
suggestive of a propagating wave.
\citet{wan15} found the first AIA wave event in agreement with a fundamental standing slow 
mode wave, and clear evidence for thermal conduction suppression in the hot 
($\gtrsim$9 MK) flare loop by coronal seismology. They suggested that compressive viscosity 
dominates in the wave dissipation. 

It is known that the classical Spitzer form of conductivity is valid under the assumptions that 
the electron velocity distribution is locally close to Maxwellian and the mean free path 
$\lambda$ is much smaller than the temperature gradient scale length $L_T$ \citep{ros86}. 
Laboratory experiments and numerical studies showed that the actual conductivity is smaller 
(by at least a factor of two) than that given by Spitzer when $L_T\lesssim30\lambda$
\citep[e.g.][]{luc83}. This is the case for typical solar flare loops with higher temperature
($T$) because $\lambda$ increases with $T^2$. For example, for hot ($T$=10 MK) loops with the length
$L$=10 Mm and electron number density $n=10^{10}$ cm$^{-3}$ (or $L$=100 Mm and $n=10^9$ cm$^{-3}$),
we find $L_T/\lambda\approx 7(L/10\,{\rm Mm})(n/10^{10}\,{\rm cm}^{-3})/(T/10\,{\rm MK})^2<30$.
For a long-duration event (LDE) with the slower-than-expected decay rate in soft X-ray (SXR),
\citet{mct93} suggested that the long duration was caused by either continuous heating (after
the impulsive burst in hard X-ray) or thermal conduction suppression. By studying the evolution 
of the SXR loop-top sources, \citet{jia06} showed that plasma waves or turbulence may play an
important role in suppressing the conduction during the decay phase of flares. 

The study presented here is a supplement to \citet{wan15} (thereafter, called Paper I). 
We determine the polytropic index directly based 
on the polytropic assumption to examine the possible effect of nonlinearity on measurements.
We also explore the effect of conduction suppression on the flare loop cooling, 
and discuss the possible cause. 

\begin{figure}[]
\begin{center}
 \includegraphics[width=5.in]{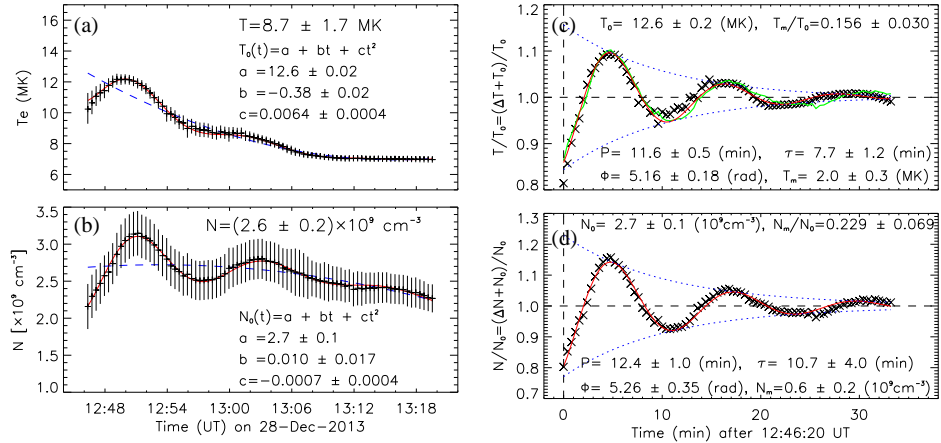} 
 \vspace*{-0.2 cm}
 \caption{Evolution of the average temperature (a) and electron density (b) for the small
loop segment shown in Figure\,\ref{fig1}(a). The solid curves show the best fit to 
the oscillatory signals, and the dashed curve is the parabolic trend. The error bars are the 
1$\sigma$ error for the Gaussian fits of DEM. $T_0(t)$ and $N_0(t)$ marked on the plots are the
fitted background trends for loop temperature and density. (c) Time profile of the temperature 
({\it crosses}) and its best fit ({\it thin solid curve}) normalized to the background trend. 
(d) Same as (c) but for electron density. The measured physical parameters of the waves are 
marked on the plots. The thick solid curve in (c) is the expected variation in temperature 
derived from the observed density variation $N/N_0(t)$ for an adiabatic process.}
   \label{fig2}
\end{center}
\end{figure}

\section{Observations and Results}
We analyzed the loop dynamics and thermal property using the SDO/AIA data. Figure\,\ref{fig1}(a)
shows that a longitudinal wave event was triggered in a large hot loop by a C-class flare near
the footpoint seen in the 131 \AA\ channel ($\sim$11 MK). The loop oscillations displayed
as alternate brightenings in the two opposite legs, which can be obviously seen 
in a time-distance plot and the animations in Paper I. The oscillation period ($P$) 
is about 12 min. The loop length ($L$) is an
important parameter for identifying the wave mode. We determine the loop 3D geometry using 
the curvature radius maximization method which assumes the line-of-sight (LOS)
coordinates (Z) of the observed loop to be same as those of a circular model \citep{asc09}.
Figure\,\ref{fig1} shows the solution of the 3D reconstruction (which has an identical 
2D projection as the observed loop), with de-projections into the XZ- and YZ-planes.
We obtained the loop length $L\simeq$180 Mm, and an estimate of the wave phase speed
$V_p=2L/P\simeq$500 km~s$^{-1}$. The phase speed is close to the sound speed of
480 km~s$^{-1}$ for the hot loop of $\sim$10 MK supporting the interpretation of the observed 
waves as a fundamental standing slow mode.

\begin{figure}[]
\begin{center}
 \includegraphics[width=4in]{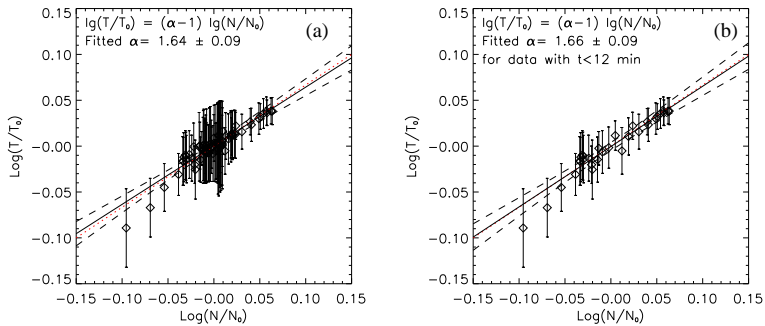} 
 \vspace*{-0.4 cm}
 \caption{Measurements of the polytropic index ($\alpha$). (a) The scatter plot of the electron 
density and temperature variations normalized to the background trend ({\it pluses}) in logarithm, 
with the linear fit ({\it solid line}) 
and the line of $\gamma$=5/3 ({\it dotted}). The dashed lines indicate the $\pm$1$\sigma$ 
fitting error. (b) Same as (a) but for the data during the initial 12-min time.  
The measured values of $\alpha$ are marked on the plots.}
   \label{fig3}
\end{center}
\end{figure}

We utilized a regularized differential emission measure (DEM) analysis on AIA images in 
six extreme-ultraviolet (EUV) bands to diagnose the temperature and electron density 
of the oscillating loop
\citep{han13}. In inversions we took a 10\% uncertainty for the 94 \AA\ and
131 \AA\ channels while a 20\% uncertainty for the other channels because the oscillations
were mainly seen in the hot channels. Figure\,\ref{fig1}(d) shows the derived DEM profile for
a small segment located at the brightest part of the loop. By assuming that the hottest
component in triple-Gaussian fits came from the flare loop, we determined the temperature
and electron density of the oscillating loop \citep[see the details in Paper I and][]{sun13}.
Figures\,\ref{fig2}(a) and (b) show their temporal variations. By fitting to a damped 
sine-function with a parabolic trend, we measured the physical parameters of the wave and
loop background plasma which are marked on the plots. 

We found that the temperature and density oscillations have similar 
periods and they are nearly in phase (Figs.\,\ref{fig2}(c) and (d)). The phase shift measured
using the cross correlation is about 12$^{\circ}$ which corresponds to the data cadence of
24 s. This nearly inphase relationship may suggests an adiabatic process on the timescale of oscillations because otherwise a large phase shift between the temperature and density 
oscillations caused by the nonideal effects such as conductive loss at higher temperature
plasma would be expected (see Paper I). We calculated the expected variation in
temperature ($T/T_0(t)$) during an adiabatic process from the measured density variation ($N/N_0(t)$)
using $T/T_0=(N/N_0)^{\gamma-1}$ (assuming the adiabatic index $\gamma$=5/3 for fully ionized
coronal plasma). We found that the expected and observed variations are in good agreement 
(except for the near-ending time of 10 min, see Fig.\,\ref{fig2}(c)). 
We quantitatively measured the polytropic index $\alpha$ under
the polytropic assumption using the following linear relationship,
\begin{equation}
{\rm log}\left(\frac{T}{T_0}\right)\,=\,(\alpha-1)~{\rm log}\left(\frac{N}{N_0}\right). \label{eqalp}
\end{equation}
This method is distinct from that used in Paper I where a linear approximation was
made. Note that to correctly apply Eq.\,(\ref{eqalp}) to measure $\alpha$ the phase shift between
$T/T_0$ and $N/N_0$ (if non-negligible) should be first removed. Figure\,\ref{fig3} shows
the linear-squares fitting for measurements of $\alpha$ in the two cases: $\alpha=1.64\pm0.09$ 
for all the data and $\alpha=1.66\pm0.09$ for only the data with $t<12$ min, where the uncertainty 
is the 1$\sigma$ error from the fit. We found that the measured values for $\alpha$ are same as
those in Paper I. This indicates that the effect of nonlinearity is negligible,
consistent with the signature that the temperature and density oscillations follow well the (damped)
sinusoidal wave. 

\begin{figure}[]
\begin{center}
 \includegraphics[width=4.8in]{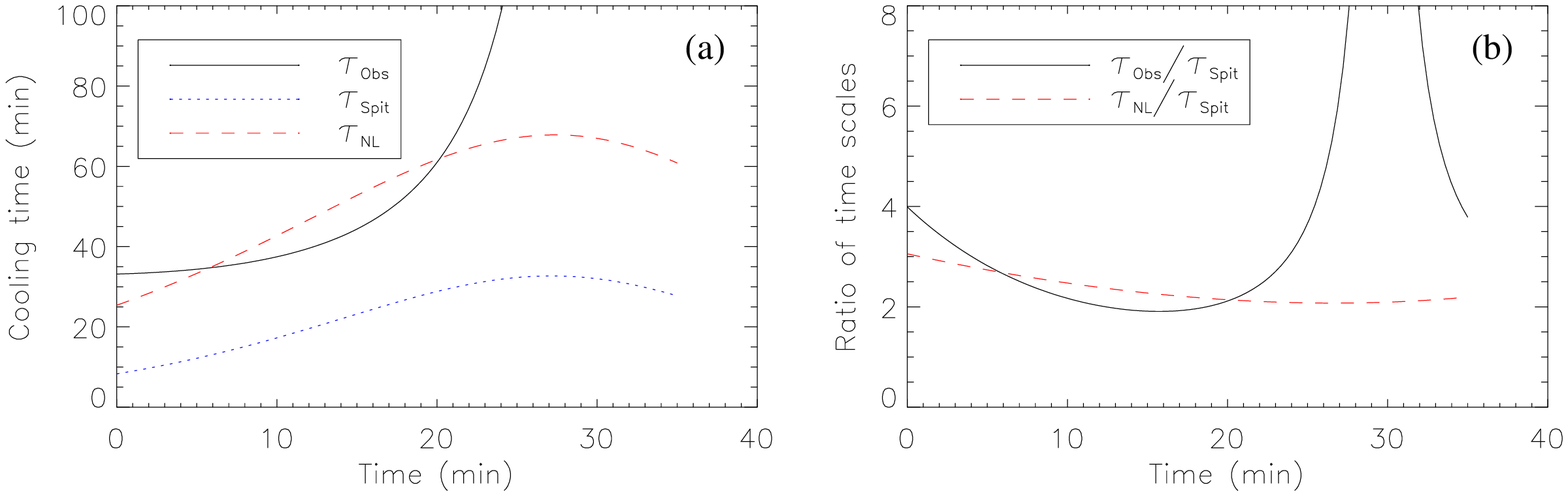} 
 \vspace*{-0.2 cm}
 \caption{(a) The cooling times estimated from the observed temperature evolution 
({\it solid line}), Spitzer conduction ({\it dotted line}), and nonlocal conduction approximation
({\it dashed line}). (b) The ratios of the measured cooling timescale to that of
Spizter ({\it solid line}), and to that of nonlocal conduction ({\it dashed line}).}
   \label{fig4}
\end{center}
\end{figure}

\section{Discussion and Conclusions}
\label{sct3}

We studied a longitudinal wave event triggered by the non-eruptive flare using SDO/AIA. 
The waves in the hot flaring loop were identified as a fundamental standing slow mode. 
We analyzed the plasma thermal and wave properties of the oscillating loop, 
and found that its temperature and electron density
variations are nearly in phase and the measured polytropic index $\alpha$ agrees well
with the adiabatic index of 5/3 for a fully ionized ideal plasma. These results imply
that the MHD energy equation can be well represented with an adiabatic form, 
or the nonideal effects such as the stratification, optically thin radiative loss, 
and heat conduction are negligible during the oscillation period. 
In Paper I, based on a 1D linear MHD model of slow waves, 
we argued that because thermal conduction dominates in the energy equation 
in the hot ($\gtrsim$9 MK) plasma, the interpretation suggests a significant reduction 
of thermal conductivity (by at least a factor of 3 as estimated quantitatively). 

The dissipation of slow waves by thermal conduction is due to temperature variations
along the loop caused by the wave, while thermal conduction causes the hot loop cooling
due to its rooting on the cool chromosphere. Now that thermal conduction is suppressed
as known from the wave analysis, its influence on the loop cooling would also
be expected. Figure\,\ref{fig4} shows that the observed
cooling timescales (calculated by $\tau_{\rm obs}=T/(dT/dt)$) is about a factor 2--4 longer
than the cooling timescales based on Spitzer's thermal conductivity 
\citep[$\tau_{\rm Spit}$ using Eq.\,(D2) in][]{sun13}.
This slower-than-expected cooling rate can be explained by conduction suppression which
has a suppression factor consistent with that inferred from the wave analysis.
For the mean temperature and density, we obtained $L_T/\lambda\sim40$ where $L_T$ is taken
as the loop length $L$. Under the nonlocal
conduction approximation \citep{jia06}, we calculated the modified Spitzer cooling 
timescales by $\tau_{\rm NL}\simeq{9.1}(L_T/\lambda)^{-0.36}\tau_{\rm Spit}$. 
Figure\,\ref{fig4} shows a good agreement between $\tau_{\rm NL}$ and $\tau_{\rm obs}$,
suggesting that the nonlocal conduction in hot plasmas may account for the observed conduction 
suppression in this event. 

In addition, the thermal conduction suppression mechanism may be used to explain the phenomenon
that the larger (solar and stellar) flares tend to be hotter than expected by the EM-$T$ relation
where $T$ is the peak temperature and EM the volume emission measure \citep{fel95, shi02}.
Assuming the balance between conduction cooling and reconnection heating and the pressure
balance for flare loops, \citet{shi02} derived the scaling law ${\rm EM}\propto{B}^{-5}T^{17/2}$
where $B$ is the magnetic field strength. If considering the suppressed conductivity
$\kappa_S=\kappa_0/S$ where $\kappa_0\simeq10^{-6}$ cgs is the classical Spitzer conductivity and
$S$ the suppression factor, we obtained the modified scaling law 
${\rm EM}_S\propto S^{-3}{B}^{-5}T^{17/2}$ as well as the relations $T_S/T=S^{6/17}$ and 
$B_S/B=1/S^{3/5}$. Given $S$=3, for instance, we estimated that the conduction suppression 
would lead to the flare loop hotter by about a factor 1.5; if the conduction suppression effect
is neglected, the magnetic field strength of stellar flares may be overestimated by
a factor of 2 from fitting the observed EM-$T$ diagram.

\vspace{0.3 cm}
The work of TW and LO was supported by NASA grant NNX12AB34G and the NASA Cooperative Agreement
NNG11PL10A to CUA. EP thanks the support from NASA grant NNX12AB34G.
SDO is a mission for NASA’s Living With a Star (LWS) program.

\end{document}